\documentclass[a4paper, reprint, twocolumn, amssymb, aps, prl, groupedaddress, superscriptaddress, nofootinbib]{revtex4-2}
\usepackage{graphicx}
\usepackage{epsf}
\usepackage{bm}
\usepackage{amsmath}
\usepackage{amsfonts}
\usepackage{amssymb}
\usepackage{epstopdf}
\usepackage{natbib}
\usepackage{color}
\usepackage[dvipsnames]{xcolor}
\usepackage{verbatim}
\usepackage{multirow}
\usepackage{bm}
\usepackage{microtype}
\usepackage[capitalize]{cleveref}
\usepackage[normalem]{ulem}

\makeatletter\let\expandableinput\@@input\makeatother
\begin{document}
    
\title{\boldmath The Nature of Phantom Dark Energy and its Relation to Time Crystals}







\author{Laura Mersini-Houghton}
\affiliation{Department of Physics and Astronomy, UNC-Chapel Hill,
NC, USA.}

\email{mersini@physics.unc.edu}
\hbadness=99999
\begin{abstract}
In this letter we show that, at its fundamental level, phantom dark energy is a fluid made up of time crystals that permeate the fabric of space-time. In turn, any well-behaved stable theory of a non-canonical scalar field that acts as phantom dark energy also leads to the production of time crystals. This relation offers a new way of thinking and a deeper insight into the nature of the most inexplicable of all ingredients in the universe, dark energy. 
\end{abstract}

\maketitle
\flushbottom


\noindent\textit{Introduction - } Dark energy is unquestionably the most mystifying energy ingredient in the universe. While many experiments are designed to measure its time evolution, and many theoretical models have been devoted to understanding dark energy, its nature at the fundamental level, remains obscure.

Previously, in ~\cite{paper1} we investigated a particular model ~\cite{Shapere2012} of time crystals from a cosmological point of view and showed that it produced phantom dark energy in the universe.

In this letter, we investigate the general case, namely whether \textit{all} phantom dark-energy models are made up of time crystals and, in turn, if a fluid of time crystals always behaves as phantom dark-energy. We claim that this relation between phantom dark energy and time crystals holds in general.

If this is indeed the case, it would shed light on the nature of phantom dark energy and imply that fundamentally it is an ideal fluid of time crystals interwoven in the fabric of spacetime. 

\noindent\textit{Phantom Dark Energy -} To prove this claim, we first investigate the conditions on a scalar field required by phantom dark energy.

Let us begin with a generalized Lagrangian
\begin{equation}
    \mathcal{L} = f(\phi) g(X) - V(\phi),
    \label{eq:lagrangian}
\end{equation}

where $f(\phi)$, $V(\phi)$ are positive semi-definite functions of the scalar field $\phi$; $X = \frac{\dot{\phi}^2}{2}$, dot is the time derivative, and \(g(X)\) is a function of $X$

We can recover the canonical case of an ideal fluid by taking \(f(\phi)=1, g(X) =X\). However, we show next that canonical Lagrangians cannot produce phantom dark energy.

For the ideal fluid given by Eqn.~\eqref{eq:lagrangian}, the pressure of the fluid \(p\) is given by the Lagrangian \(\mathcal{L} = p\) of the scalar field, and its energy density \(\rho\) by the respective Hamiltonian of the field, \( \mathcal{H} = \rho\).
The transformation between the Lagrangian and the Hamiltonian is obtained by:
\begin{equation}
\mathcal{H} + \mathcal{L} = \Pi \dot{\phi}
\end{equation}
where the conjugate momentum is
\begin{equation}
\Pi= \frac{\partial\mathcal{L}}{\partial \dot{\phi}} = f(\phi) g'(X) \dot{\phi}\, .
\label{eq:eqmomentum}
\end{equation}

Therefore,
\begin{equation}
\mathcal{H} = f(\phi) ( 2 X g'(X) - g(X) ) + V(\phi)
\label{eq:hamiltonian}
\end{equation}

with the Lagrangian of the scalar field $\mathcal{L}$ given by Eqn.~\eqref{eq:lagrangian}, where prime denotes $\frac{\partial}{\partial X}$. 
We require the energy of the scalar field given by the Hamiltonian of Eqn.~\eqref{eq:hamiltonian} to be positive, \(\rho>0\).

Phantom behavior demands the equation of state of the ideal fluid to be: \(w_{\phi} \leq - 1\), where

\begin{equation}
    w_{\phi} =\frac{p}{\rho} = \frac{f(\phi) g(X) -V(\phi)}{f(\phi) (2X g'(X) - g(X) ) +V(\phi)}
    \label{eq:eos}    
\end{equation}
which is equivalent to demanding \(\rho + p \le 0\).

Since \(\rho + p= f(\phi) 2 X g'(X)\), a phantom dark energy fluid that satisfies \(w_{\phi} \leq -1\) leads to condition \(g'(X) < 0 \) given that \(X \geq 0\).

The latter implies that for a phantom dark energy fluid, the function \(g(X)\) must be a decreasing function of \(X\), which proves that \(g(X)\) cannot belong to the class of canonical Lagrangians that are linear in \(X\), i.e., \( g(X) \ne X\).

Furthermore, the special case of \(g(X)\) which is linear in \(X\) but with the 'wrong' negative sign, \(g(X)= - \kappa X\), where \(\kappa>0\) is a parameter, is also excluded because it is unbounded from below leading to unbounded negative energies \cite{wands, Wilczek2012,Shapere2012}, which cannot give rise to a Friedmann-Roberson-Walker (FRW) universe. (It is straightforward to see that a rotation of the field \(\phi -> i\psi\) would recover the canonical case \(g(X) = \kappa \dot{\psi}^{2}/2\), that we ruled out as a candidate for phantom dark-energy models.)

Einstein field equations relate the expansion of the universe to the dark energy fluid pressure \(p\) and energy density \(\rho\) such that, modulo prefactors of \(8\pi G/3\), the Hubble parameter is \(H^{2} =(\frac{\dot{a}}{a})^2 \simeq \rho\) and \(\dot{H} \simeq - (\rho+p)\), relations that rule out scalar field models with unbounded negative energies of the field discussed above. 
 
The Bianchi identity \(\dot{\rho} + 3H (\rho +p)=0\) is a combination of the above two Einstein's equations. It is equivalent to the field's equation of motion, namely:

\begin{equation}
    \dot{\phi} [ \dot{\Pi} + 3 H \Pi + \frac{\partial V(\phi)}{\partial\phi} - g(X) \frac{\partial f(\phi)}{\partial \phi}] = 0
\label{eq:eom}
\end{equation}

 with \(\Pi\) the conjugate momentum of the field in Eqn.~\eqref{eq:eqmomentum}. This equation can be satisfied by the trivial solution \(\dot{\phi}=0\), or when the squared bracket equals zero, that is, by solving the equation for conservation of momentum, the function \(\Pi(X)\). Given a knowledge of \(V(\phi)\) and \(f(\phi)\), Eqn.~\eqref{eq:eom} can be solved numerically.
 However, it should be noted that for phantom dark energy the Hubble parameter grows rapidly with time and soon overwhelms the last two unspecified potential energy terms, a point that will be important to discussing the role of potential terms below.

 This set of equations, along with the requirement that its energy density be positive, place constraints on the phantom dark energy stress energy tensor. (In principle, we could include other matter and radiation contributions to the above stress energy components that enter Einstein's equations, however as is well known \cite{caldwelldoomsday, caldwell} those components dilute away fast and phantom dark energy dominates the expansion.)

For these reasons, well-behaved stable phantom dark energy models belong to the class of non-canonical Lagrangians which, in addition to the linear term, contain higher-order terms in powers of \(X\), meaning \(g''(X) \ne 0\). In fact, since the requirement for phantom behavior \(w_{\phi} \le -1\) leads to \(g'(X) \le 0\) that condition places a constraint on the second derivative \(g''(X)\ge 0\) for the energy to be bound from below and for the theory to be stable.

\noindent\textit{Time Crystals - } As shown above, the class of phantom dark energy models with the non-canonical kinetic term \(g(X)\) is such that it generically contains higher powers of \(X\) besides the linear term in \(X\), that is \(g''(X)\ne 0\). In fact, we further know that since \(g'(X) < 0\) for \(w_{\phi} < -1\) then \(g''(X) >0\). Hence, the Hamiltonian \(\mathcal{H}\) of Eqn.~\eqref{eq:hamiltonian} which provides the energy density of the non-canonical fluid, obtained from the Lagrangian \(\mathcal{L}\) will be related to the Lagrangian in a non-smooth way, given the above conditions on the function \(g(X)\), and the expression of conjugate momentum, Eqn.~\eqref{eq:eqmomentum},
\(\Pi = \dot{\phi} g'(X) f(\phi)\). 

Since \(\Pi\) is a nonlinear function of \(\dot{\phi}\)  and of \(g(X)\) which has linear and quadratic terms of opposite signs, (and possibly higher order powers of \(X\)), then it can be seen from the expression of momentum Eqn.~\eqref{eq:eqmomentum} and of the Hamiltonian Eqn.~\eqref{eq:hamiltonian}, that \(\mathcal{H}\) will be a multi-valued function of (\(\dot{\phi}\), \(\phi\)) indicating the existence of orbits, with solutions \(\dot{\phi}\ne 0\), along which the kinetic energy is minimized.

Let us show explicitly the existence of orbits and highlight the fact that the non-smooth relation between the Hamiltonian and the Lagrangian which is a consequence of the constrained non-canonical kinetic function \(g(X)\), breaks the time translation symmetry, i.e. these turning points bounding the orbit of the crystal are found at \(\dot{\phi}_{t} \ne 0\).

Kinetic energy is minimized at
\begin{equation}
    \frac{\partial \mathcal{H}}{\partial \dot{\phi}} = \frac{\partial \mathcal{H}}{\partial X} = 0\,.
    \label{eq:keom}
    \end{equation}
This equation is equivalent to finding cusps of conjugate momentum \(\Pi\)
    \begin{equation}    
    \frac{\partial \Pi}{\partial \dot{\phi}} =0 \,
    \label{eq:momenta}
\end{equation}
\text{at} \(\dot{\phi} = \dot{\phi}_{t} = \pm \sqrt{X_{t}} \).
They result in the following constraint equation at \(\dot{\phi} =\dot{\phi}_t\):
\begin{equation}
g'(X) + 2 X g''(X) = 0\,
\label{eq:constrainong}
\end{equation}

The turning points \(\phi_{t}, X_t\) are found by solving Eqn.~\eqref{eq:constrainong}. Functions \(g(X)\) and \(\Pi\) are non linear functions of \(\dot{\phi}\) with the first term \(g' <0 \) and the second term \(g'' >0\) in Eqn.~\eqref{eq:constrainong}. Hence, due to the constraints on the nonlinear structure of the function \(g(X)\), solutions given by the cusps of the momentum function which minimize energy will be found at nontrivial points \(\dot{\phi} \ne 0\) thereby indicating motion of the bound state that breaks time translation symmetry, yielding a time crystal solution. 

We came across a specific example of time crystals producing phantom dark energy in \cite{paper1}. Another illustration that satisfies the conditions on \(g'(X), g''(X)\) would be a kinetic term of the form \(g(X)= e^{- \alpha X}\) which, according to Eqn.~\eqref{eq:constrainong} gives the turning points at \(\dot{\phi}_{t} = \pm \sqrt{X_{t}}= \pm \sqrt{1/2\alpha}\). In general we can expand any nonlinear function as a series expansion \(g(X)= \Sigma \alpha_{n} X^n\) for \(n =1,2,3,4...\) with coefficients \(\alpha_n\) constrained by the requirements \(g'(X)<0 , g''(X)>0\) and obtain the turning points from the cusps of the momentum. In this case, the solutions to Eqn.~\eqref{eq:constrainong} are \(X_{t} = |\alpha_{n}|/2(n+1)|\alpha_{n+1}| \ne 0 \). 

In the general case, independently of the potential \(V(\phi)\), solutions of Eqn.~\eqref{eq:constrainong} at the cusps of the momentum with \(\dot{\phi}\ne 0\) are bound states that minimize kinetic energy bound by the turning points at \(\dot{\phi} = \pm \dot{\phi_t} = \pm \sqrt{2X_t} \ne 0\) of the time crystal. In sum, the same conditions placed on phantom dark energy require a non-canonical multi-valued Hamiltonian \(\mathcal{H}\) and nonlinear expression of the momentum \(\Pi \simeq \dot{\phi}g'(X)\) also imply that the minima of these expressions will also occur at nontrivial points \(\dot{\phi_t} \neq 0\). The fact that the cusps of the conjugate momentum \(\Pi\) for this multi-valued energy function \(\mathcal{H}\) of \(\dot{\phi}\) correspond to the boundary of the time crystal structure in field space becomes apparent from noticing that there is motion \(\dot{\phi} \neq 0\) of the crystal as a whole that breaks the time translation symmetry.

We have seen until now that the potential energy \(V(\phi)\) has not played an important role in our discussion of phantom dark energy or time crystals since orbits of the time crystal are found by minimizing kinetic terms. Yet, the Bianchi identity which give momentum conservation Eqn.~\eqref{eq:eom} depends on the potential energy term. Here we would like to understand whether our results are completely general for any choice of potential energy or whether the potential needs to be fine tuned to specific cases. (For the sake of simplicity, we explore this question by setting \(f(\phi)=1\) without losing generality, it can always be factored out by redefining the potential \(\tilde{V}(\phi) = V(\phi)/f(\phi)\).)  

From the Lagrangian equation of motion or equivalently the Bianchi identity \(\dot{\Pi}= \frac{\partial\mathcal{L}}{\partial \phi} - 3 H \Pi \) which simplifies to \(\dot{\Pi}  + \frac{\partial V(\phi)}{\partial \phi} + 3 H \Pi=0\), where \(H\) is the Hubble parameter. In a phantom dark energy universe, the Hubble parameter grows with time as \(H\simeq a(t)^{-\frac{3}{2} (1+w_{\phi})}\), where \((1+w_{\phi})\le 0\) and \(a(t)\) is the scale factor. It quickly dominates over the potential energy term \(\frac{\frac{\partial V(\phi)}{\partial\phi}}{3H} \rightarrow 0\) for any reasonable potential rendering the potential term insignificant, effectively driving it to a minimum, in the equation of motion.  In this sense, the scalar field enters an attractor behavior due to the Hubble drag term, (see ~\cite{ArmendarizPicon2001} for further discussion.) 

The term 'reasonable potentials' is used here to denote potentials \(V(\phi), f(\phi)\) that satisfy the positive energy condition \(\rho \ge 0)\), which is achieved for \( g(X_{t}) \le \frac{V(\phi)}{f(\phi)}\).  The latter restriction on the class of potentials is very weak since the kinetic function \(g(X)\) is bound to take values within the time crystal region at \(X\le X_t\) and one can always rescale the potential by an overall constant factor to satisfy the inequality. 

Therefore, independently of the chosen functional form of the potential, and owing to a fast growing Hubble drag term which overwhelms the potential term, in general, non canonical time crystal models, with orbits that minimize kinetic energy and have motion \(\dot{\phi} \ne 0\) which breaks the underlying symmetry, will produce phantom dark energy. 

\noindent\textit{Stability of Time Crystals - } We need to show that the region of time crystals bounded by the turning points, given by the cusps of the conjugate momentum, corresponds to stable bound states.

Before time crystals were discovered, authors of \cite{wands} who studied this class of Lagrangians in the context of k-essence naturally interpreted the boundary of the time crystal with a terminating singularity since, as we will see next, the speed of sound diverges there. They relied on Einstein equations, to conclude from Friedman's equation and the diverging speed of sound at the boundary, that the expansion of the universe terminates abruptly when the field reaches its boundary value \(\phi_t\). 

In fact, as shown a few years later in \cite{Wilczek2012, Shapere2012} when time crystal solutions were discovered, time crystals are stable bound state solutions and their boundary at \(\phi=\phi_t\) behaves like a brick wall potential for the field, namely as the field approaches its boundary, the speed of sound squared diverges, implying that the field simply reverses its momentum there while all the time conserving the energy thus preserving the bound state. In this work, the phantom dark energy fluid is collections of these bound states, of the time crystals, and we show next that the expansion of the universe does not suffer a terminating singularity. Rather, it continues into an accelerated expansion until doomsday: the Big Rip.

Let's estimate the speed of sound squared and show that \(C_s^{2} >0\) everywhere in the time crystal field region with \(\phi \leq \phi_t\) and it diverges at the boundary \(\phi_t\). In other words, the bound state structure is stable against small perturbations and the field cannot break the crystal structure to jump over values higher than its orbit \(\dot{\phi_t}\) without investing a large amount of work and energy to break the crystal-bound state, because at the boundary the speed of sound squared \(C_s^2 -> \infty\) at \(\dot{\phi} -> \dot{\phi_t}\). At the boundary, the field bounces back by reversing its momentum and remains confined within the well-behaved bound state solutions. The diverging speed of sound at the boundary simply serves to reinforce the stability of the crystal by keeping the field bound within the crystal regime despite the accelerated expansion of the universe, discussed in the context of the field's equation of motion above.

Actually, since the fluid is a collection of bound state crystals, we expect to find the speed of sound to be not only positive but to be \(C_{s}^{2} \ge 1\) characteristic of the crystalline structures. If such a speed of sound is found observationally, it would be an indicator of a crystalline type of dark energy. The relation of the speed of sound to background perturbations, implies that gravitational collapse of background perturbations into dark matter will be highly dampened, and none of the symmetries, such as Lorentz invariance, are violated,  as explained in detail in ~\cite{ArmendarizPicon2001, caldwell}.

The sound speed squared \( C_s^2 \) for an ideal fluid with a stress energy tensor
\begin{equation}
T_{ab} = \rho u_{a} u_{b} + g_{ab}(\rho + p)
\end{equation}
with \(u_a\) the four vector normalized to unity and \(g_{ab}\) the metric of the FRW universe, is defined as:
\begin{equation}
    C_s^2 = \frac{\partial p}{\partial \rho} = \frac{\mathcal{L'}_X}{\mathcal{H'}_X}.
\label{eq:speedofsound}
\end{equation}
where the subscript \(X\) indicates derivatives with respect to that variable. Given the Lagrangian and Hamiltonian expressions in Eqn.~\eqref{eq:lagrangian} and Eqn.~\eqref{eq:hamiltonian}, in the general non-canonical case the speed of sound is:

\begin{equation}
    C_s^{2} = \frac{g'(X)}{ ( g'(X) + 2 X g" (X) )} 
    \label{eq:noncanonicalspeed}
\end{equation}
The stability condition requires that the speed of sound is positive, \(C_{s}^{2}>0\). The latter, in combination with the condition from the phantom equation of state \(g'(X)\le 0\), and from Eqn. ~\eqref{eq:noncanonicalspeed} implies that \(g'(X) + 2X g''(X) \le 0\). Given the stability condition of the fluid, \(g''(X) >0\), and phantom behavior \(g'(X) <0 \) naturally results not only in the speed of sounds being positive but, furthermore, in \(C_{s}^{2} \ge 1\), as can be seen from Eqn. ~\eqref{eq:noncanonicalspeed}.

Note from the Hamilton equations, Eqn. \eqref{eq:eom}, that the exact same condition that minimizes kinetic energy \(\mathcal{H'}_X =0 \)  at \(\dot{\phi} = \pm \dot{\phi}_t\) and gives the turning points of the time crystals, also enters in the denominator of the expression for the speed of sound squared, Eqn.~\eqref{eq:speedofsound}. Therefore, given that the nominator \(\mathcal{L'}_{X} < 0 \) for the phantom dark energy fluid ( \(g'(X) < 0\) in Eqn.~\eqref{eq:noncanonicalspeed} ), it can be clearly seen that the speed of sound squared will diverge at the boundary as expected
\begin{equation} 
C_s^2 =\frac{\mathcal{L'}_X}{\mathcal{H'}_X} =  \infty\quad \textit{at}\quad\dot{\phi} =\pm \dot{\phi}_t
\end{equation}
since \(\mathcal{H'}_X =0\) is the requirement for minimizing the energy along the orbit and it defines the time crystal, with turning points at \(\dot{\phi}_t \ne 0\) which break the time translation symmetry, a consequence of the fact that the crystal has motion as a whole. The turning points \(\phi_t\) with the diverging speed of sound squared define a brick wall in field space, an upper bound for the dark energy fluid defined by the non-canonical field \(\phi\) which is confined to take values within its bound state, the time crystal. Energy is conserved at the turning points as the field reverses momentum, therefore the expansion of the universe does not terminate abruptly. Instead the field is forced to oscillate within the time crystal region. 

\noindent\textit{Conclusions - } In this letter we demonstrated that the conditions placed on a scalar field with a time independent Hamiltonian corresponding to the phantom dark energy fluid are equivalent to those that produce time crystals - stable bound states with motion in time \(\dot{\phi}\ne 0\) which break the underlying time symmetry of the Hamiltonian.
We argued on general grounds that the potential \(V(\phi) \) is overpowered by the Hubble drag term in the field's equation of motion and does not play an important role in the conditions of a phantom dark energy equation of state \(w_{\phi} \leq -1\) or in its speed of sound squared. This means that as long as the conditions for a phantom fluid are satisfied, time crystals that make up the phantom dark energy fluid will always exist, independently of the particular model.

The Hamiltonian equation for minimizing energy defined the time crystal orbit. These energy conditions show that the same structures which are stable along the orbit of the crystal, break time symmetry and minimize energy, with a spike \(C_s^{2} -> \infty\) at the boundary, defining a brick wall in field space where the field reverses its momentum and is contained within the stable bound state regime, but a finite \(C_s^2 \ge 0\) everywhere else within the time crystal region. Simultaneously, they give rise to the phantom dark energy fluid equation of state on the background spacetime, satisfying the conditions \(w_{\phi}<-1\) and the need for a non canonical Lagrangian for the phantom fluid.

Given the nature of phantom dark energy discussed in this letter, we expect the propagation of perturbations to have \(C_s^2 \ge 1\) for the phantom dark energy made up of time crystals, typical of crystalline structures. The latter contributes a term in the perturbation equation ~\cite{Garriga1999} which leads to the suppression of structures and dampening of the gravitational instabilities that would have produced cold dark matter, see ~\cite{caldwell}.

In conclusion: we have shown that phantom dark energy fluid produces time crystals and conversely time crystals give rise to phantom ideal fluids. The two are equivalent for generic case non-canonical models. 

 If dark energy in our universe turns out to be of a phantom type, then whenever we look at our sky we glimpse at a spacetime threaded with times crystals scintillating simultaneously throughout the universe and beaming with phantom dark energy. As the volume of the universe expands and races rapidly towards the Big Rip, every instant the fabric of our spacetime is being suffused with ever more time crystals. 

Furthermore, the unique signature on the propagation of perturbations through the speed of sound squared, \(C_{s}^2 >1 \) which are expected from crystalline structure, although it doesn't violate Lorentz invariance and causality as discussed in ~\cite{caldwell}, it can still give rise to testable observations in the CMB sky such as: shifting of the baryonic acoustic peaks; damped perturbations that contribute to cold dark matter sector; signatures on the ISW region of CMB's low multipoles, and possibly on \(\sigma_8\). We explore these possibilities in a separate work.

In this letter we offer a new way of thinking about the nature of phantom dark energy by revealing its connection to exotic crystalline structures - time crystals. Perhaps, similar relations with crystalline structures may be found in the future for other forms of dark energy, including a pure vacuum energy.

\noindent\textit{Acknowledgment:}
L.Mersini-Houghton is grateful to O.Akarsu, E. Di Valentino, and J.Ng for useful discussions and to Klingsberg foundation and the Bahnson fund for their support.





\begin{thebibliography}{99}

\bibitem{ArmendarizPicon2000}
C.~Armendariz-Picon, V.~F.~Mukhanov, and P.~J.~Steinhardt,
\emph{A Dynamical solution to the problem of a small cosmological constant and late time cosmic acceleration,}
Phys.\ Rev.\ Lett.\ \textbf{85}, 4438 (2000) ,
[arXiv:astro-ph/0004134].

\bibitem{Chiba2000}
T.~Chiba, T.~Okabe, and M.~Yamaguchi,
\emph{Kinetically driven quintessence},
Phys.\ Rev.\ D \textbf{62}, 023511 (2000)
[arXiv:astro-ph/9912463].

\bibitem{ArmendarizPicon2001}
C.~Armendariz-Picon, V.~F.~Mukhanov, and P.~J.~Steinhardt,
\emph{Essentials of k essence},
Phys.\ Rev.\ D \textbf{63}, 103510 (2001)
[arXiv:astro-ph/0006373].

\bibitem{Garriga1999}
J.~Garriga and V.~F.~Mukhanov,
\emph{Perturbations in k-inflation},
Phys.\ Lett.\ B \textbf{458}, 219 (1999)
[arXiv:hep-th/9904176].

\bibitem{Melchiorri_2003}
A.~Melchiorri, L. ~Mersini-Houghton, C.~J.~Ödman, and M.~Trodden
\emph{The State of the Dark Energy Equation of State}
Phys. \ Rev. \ D \textbf{68}, 043509 (2003),
[arXiv:astro-ph/0211522].

\bibitem{Shapere2012}
A.~Shapere and F.~Wilczek,
\emph{Classical Time Crystals},
Phys.\ Rev.\ Lett.\ \textbf{109}, 160402 (2012)
[arXiv:1202.2537 [cond-mat.other]].

\bibitem{Wilczek2012}
F.~Wilczek,
\emph{Quantum Time Crystals},
Phys.\ Rev.\ Lett.\ \textbf{109}, 160401 (2012)
[arXiv:1202.2539 [quant-ph]].

\bibitem{timecrystalreview}
K.~Sacha and J.~Zakrzewski
\emph{Time Crystals: A Review}
Rep. \ Prog. \ Phys. \textbf{81}  016401 (2018)
[arXiv:quant-ph 1704.03735].

\bibitem{caldwell}
J.~K.~Erickson, R.~R.~Caldwell, P.~J.~Steinhardt, C.~ Armendariz-Picon, V.~Mukhanov
\emph{Measuring the Speed of Sound of Quintessence}
Phys. \ Rev. \ Lett. \textbf{88} (2002)
[arXiv: astro-ph/0112438].


\bibitem{Hlozek2015}
R.~Hlozek \textit{et al.},
Phys.\ Rev.\ D \textbf{91}, 103512 (2015)
[arXiv:1411.1074 [astro-ph.CO]].

\bibitem{paper1}
L.~Mersini-Houghton 
\emph{Dark Energy from Time Crystals}
\emph{with referee} (2024)

\bibitem{caldwelldoomsday}
R.~Caldwell, M.~Kamionkowsky, N.~Weinberg
\emph{Phantom Energy and Cosmic Doomsday},
Phys.\ Rev.\ Lett. 91 071301 (2003), 
[arxiv: [astro-ph/0302506]]

\bibitem{wands}
P.~Jorge, J.~Mimoso, D.~Wands
\emph{On the dynamics of k-essence models},
J. \ Phys. \ Conf.\ Ser. \textbf{66}, (2006).

 \bibitem{phantom}
 Spergel.~D.~N 
 \emph{The dark side of cosmology:dark matter and dark energy}, 
 Science \textbf{347}, (2016);
 R.~J.~ Scherrer
 Phys. \ Rev. \ D. 71 (2004).

\end{thebibliography}


\end{document}